# QUANTUM TECHNOLOGY:
# THE SECOND QUANTUM REVOLUTION.

Jonathan P. Dowling[1].
*Quantum Computing Technologies Group, Section 367,*
*Jet Propulsion Laboratory,*
*Pasadena, California 91109, USA.*

Gerard J. Milburn[2],
*Department of Applied Mathematics and Theoretical Physics,*
*University of Cambridge, Wilberforce Road, Cambridge, UK.*
*and*
*Centre for Quantum Computer Technology.*
*The University of Queensland*
*St Lucia, QLD 4072, Australia;*

**Abstract.**

We are currently in the midst of a *second quantum revolution*. The first quantum revolution gave us new rules that govern physical reality. The second quantum revolution will take these rules and use them to develop new technologies. In this review we discuss the principles upon which quantum technology is based and the tools required to develop it. We discuss a number of examples of research programs that could deliver quantum technologies in coming decades including; quantum information technology, quantum electromechanical systems, coherent quantum electronics, quantum optics and coherent matter technology.

[1] Email: jonathan.p.dowling@jpl.nasa.gov

# Introduction.

We are currently in the midst of a *second quantum revolution*. The first quantum revolution gave us new rules that govern physical reality. The second quantum revolution will take these rules and use them to develop new technologies. The *First Quantum Revolution* occurred at the last turn of the century, arising out theoretical attempts to explain experiments on blackbody radiation. From that theory arose the fundamental idea of wave-particle duality—in particular the idea that matter particles sometimes behaved like waves, and that light waves sometimes acted like particles. This simple idea underlies nearly all of the scientific and technological breakthroughs associated with this *First Quantum Revolution.* Once you realize just how an electron acts like a wave, you now can understand the periodic table, chemical interactions, and electronic wavefunctions that underpin the electronic semiconductor physics. The latter technology drives the computer-chip industry and the *Information Age*. On the other hand, the realization that a light wave must be treated as a particle gives to us the under-standing we need to explain the photoelectric effect for constructing solar cells and photocopying machines. The concept of the photon is just what we need to understand the laser. By the end of this century, this first revolution of quantum mechanics has evolved into many of the core technologies underpinning modern society. However, there is a *Second Quantum Revolution* coming—which will be responsible for most of the key physical technological advances for the 21st Century.

*Quantum technology* allows us to organise and control the components of a complex system governed by the laws of quantum physics[1]. This is in contrast to conventional technology which can be understood within the framework of classical mechanics. There are two imperatives driving quantum technology. The first is practical; The dominant trend in a century of technological innovation is miniaturisation: to build devices on a smaller and smaller scale. Ultimately this will deliver devices at length scales of nanometres and action scales approaching Planck's constant. At that point design must be based on quantum principles. The second imperative is more fundamental. The principles of quantum mechanics appear to offer the promise of a vastly improved performance over what can be achieved within a classical framework.

The hallmark of this *Second Quantum Revolution* is the realization that we humans are no longer passive observers of the quantum world that Nature has given us. In the *First Quantum Revolution,* we used quantum mechanics to understand what already existed. We could explain the periodic table, but not design and build our own atoms. We could explain how metals and semiconductors behaved, but not do much to manipulate that behavior. The difference between science and technology is the ability to engineer your surroundings to your own ends, and not just explain them. In the *Second Quantum Revolutio*n, we are now actively employing quantum mechanics to alter the quantum face of our physical world. We are transforming it into highly unnatural quantum states of our own design, for our own purpose. For example, in addition to explaining the periodic table, we can make new artificial atoms—quantum dots and excitons—which we can engineer to have electronic and optical properties of our own choosing. We can create states of quantum coherent or entangled matter and energy that likely existed nowhere else in the Universe. These new man-made quantum states have novel properties of sensitivity and nonlocal correlation that have wide application to the development of computers, communications systems, sensors and compact metrological devices. Thus, although quantum mechanics as a science has matured completely, quantum engineering as a technology is now emerging on its own right. It is just a matter of being in the right place at the right time to take full advantage of these new developments.

*Quantum principles.*

The objective of quantum technology is to deliver useful devices and processes that are based on quantum principles which include:

Quantisation (quantum size effect); the allowed energies of a tightly confined system of particles are restricted to a discrete set.

- Uncertainty principle; for every perfectly specified quantum state there is always at least one measurement, the results of which are completely certain, and simultaneously at least one measurement for which the results are largely random.
- Quantum Superposition; if an event can be realized in two or more indistinguishable ways, the state of the system is a superposition of each way simultaneously.
- Tunneling; the ability of a particle to be found in spatial regions from which classical mechanics would exclude it.
- Entanglement: the superposition principle applied to certain nonlocal correlations, if a correlation can be realized in two or more indistinguishable ways, the state of the system is a superposition of all such correlations simultaneously.
- Decoherence: what happens to quantum superpositions when an attempt is made to distinguish previously indistinguishable ways an event can be realized. It renders superposi-tions of probability amplitudes into superpositions of classical probabilities. Decoherence has no analogue in classical physics.

In this review we will encounter each of these principles applied in various current and emerging technologies. While we have understood these principles for many decades, only recently has it become possible to engineer devices according to these principles. In addition to fundamental principles of quantum mechanics, quantum technology will require a set of specific tools that are generic. These include: *quantum metrology, quantum control, quantum communication and quantum computation.*

***The tools of quantum technology.***

Successful technologies are predicated on precise engineering, which in turn requires high-precision measurement. Quantum technology will thus require us to develop a *quantum metrology*. It is well known that measurement in quantum mechanics requires a radical reappraisal of traditional measurement concepts. The modern description of measurement, in terms of quantum open systems[2], suffices to develop the principles required for measurement in a technological context. While many questions remain, it is clear that quantum mechanics enables new types of high-precision measurement[3].

No complex technology can function without incorporating *control systems*; feedback, feed-forward, error correction, etc. Control systems take as input a measurement record, perform complex signal analysis in the presence of noise, and use the information gained to adapt the dynamics of the system or measurement in some way. Some years ago a simple theory of quantum control and feedback[4] was developed, which indicates that classical control theory is inadequate. The development of the general principles of quantum control theory is an essential task for a future quantum technology[5].

The recent discovery of *quantum communication* protocols[6] that are more powerful than can be achieved classically suggests new ways for components of a large complex system to be interconnected. Quantum communication technologies require new principles of operation. A *quantum internet* based on quantum optical channels, for example, would require new protocols for communication (eg. distributed quantum computing, quantum packet switching) and might incorporate quantum key distribution and error correction as fundamental components. Already it is clear that there is a need to develop an understanding of quantum communication complexity.

Quantum mechanics enables, at times, exponentially more efficient algorithms than can be implemented on a classical computer[7]. This discovery has led to the explosive growth of the field of *quantum computation.* Building a quantum computer is the greatest challenge for a future quantum technology, requiring the ability to manipulate quantum-entangled states for millions of sub-components. Such a technology will necessarily incorporate the previous three quantum applications for readout (*quantum metrology*), error correction (*quantum control*), and interconnects (*quantum communication*). A systematic development of the principles of measurement, control, and communication in a quantum world will facilitate the task of building a quantum computer.

## Quantum technologies.

### Quantum information technology.

Quantum Information Technology is the advance guard of the Second Quantum Revolution and has, in part, its origins in what were once little-known and little understood "second-order" effects predicted by the quantum theory. The earliest recognition of these effects came in 1935 in a famous paper by Einstein, Podolsky, and Rosen (EPR), who pointed out that certain carefully prepared quantum systems have nonlocal, entangled, nonclassical correlations between them[8]. These EPR correlations are not just a manifestation of the usual wave-particle duality, but rather they are a new type of higher-level quantum effect that manifests itself only in precisely engineeered man-made quantum architectures. For this reason, the first physical determination of these nonlocal EPR effects did not occur until around 1980, in small table-top quantum optical experiments by Clauser[9] and also Aspect[10].

From 1980 until 1994, the theory and experiments on nonlocal quantum correlations remained an obscure branch of the Foundations of Quantum Mechanics. However, all that changed with two breakthroughs in 1994, when two critical events took place that began the quantum information revolution. The first was the experimental demonstration, by a group at the British Defence Evaluation and Research Agency (DERA), that nonlocal photon correlations could be used to make an unbreakable quantum cryptographic key distribution system over 4 km of optical fiber[11]. The second was the theoretical exposition by Shor that a quantum computer, by harnessing these delicate nonlocal quantum entanglements, could provide an exponential speed up in computational power for some intractable numerical problems[12]. Thus, in 1994 the world learned that quantum entanglement was important technological tool and not just a curiosity. Thence, the rapidly growing fields of Quantum Information Theory and Quantum Computing were born[13]. These research areas provide the theoretical underpinnings that bind the other applied quantum technologies together; providing an over-arching theoretical and interpretive framework connecting advances in one topic to those in another.

There is now a world-wide effort to build simple quantum information processors. These devices are a long way from a general purpose quantum computer but may be very powerful special purpose machines (such as a factoring engine for code breaking). Even that outcome is a decade or more away as we need to learn how to engineer entanglement on a scale of a dozen or so particles. Many physical systems have been suggested for building a quantum computer, including: NMR, ion traps, cavity QED, quantum dots, superconducting circuits, and optical systems. (We discuss some of these below.) Each particular system will require enormous advances in the respective fields and thus hasten the development of many different quantum technologies.

**Quantum Algorithms:**

Quantum algorithms are computer programs which, when run on a quantum computer, take advantage of nonlocal, nonclassical, quantum entanglement to provide a computational advantage. In some cases this advantage is huge. In Shor's factoring algorithm, the quantum program runs exponentially faster than the best known classical program. All *Public Key Cryptography* systems—used widely on the internet—rely on the inability of computers to factor numbers rapidly. Shor's algorithm would be the "killer app" code breaker—cracking secret codes in seconds that would take trillions of years to do on a classical machine. A second quantum computer program, Grover's algorithm[14], provides a quadratic speedup over the best possible classical algorithm for searching a random database. Applications are to data mining, optimization, and code breaking. The search continues for even more programs that run on quantum machines.

**Quantum Cryptography:**

While quantum computing is still in its infancy, quantum cryptography is here and now. In fact, European and American commercial interests are in the process of developing commercial quantum key distribution systems. This technology is made possible by recent advances in single photon optical fiber engineering, which allows the distribution of quantum entangled photons over hundreds of kilometers of optical fiber, or even in air as in Earth to Space. Los Alamos National Lab (LANL) is currently working on such a freespace system in the group of Richard

Hughes. The quantum cryptographic keys distributed in this fashion are provably immune to attack—guaranteed by the *Heisenberg Uncertainty Principle*. This result, when coupled with the threat quantum computers now pose to existing public key systems, gives a new quantum way of transmitting secure data, in a fashion that is proven to be unbreakable—even by a quantum computer[15].

**Quantum Information Theory:**
In parallel to the developments in quantum computing and cryptography, there has evolved an entirely new science of quantum information theory that investigates the fundamental limits of how information can be processed in a quantum world. Due to nonlocal entanglements, which do not exist in classical information theory, all of the classical results have to be reworked to account for subtle quantum effects. There are now quantum versions of Shannon's theorem for channel capacity, as well as many other fundamental laws of a field that was hitherto only classical at its foundation. Investigations are underway in the areas of quantum data compression and superdense coding. In addition, quantum error correction codes are being developed all over the world to ensure error-free operation of a quantum computer, once the technology is in place[16]. Finally, a related field of study is that of the effects of decoherence and noise on quantum circuits and communications channels. Early on it was realized that a quantum computer might be realized in any of a number of disparate systems: ion traps, optical cavities, quantum dots, etc. It was necessary to develop a general theory of environmental degradation effects on delicate quantum systems. These new techniques are fundamental and have direct application to the production of robust quantum sensors that rely on these effects, such as the quantum optical gyro (see below).

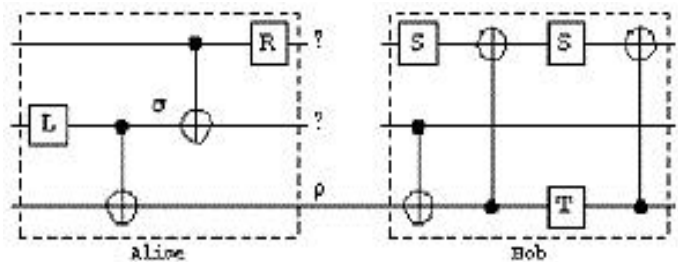

*Quantum logic circuit.*

***Quantum electromechanical systems (QEMS).***

A QEMS is a nano-fabricated mechanical system, with action of the order of Planck's constant and incorporating transducers operating at the quantum limit. Such devices enable a force microscopy sensitive enough to detect the magnetic moment of a single spin, or the deformation forces on a single macromolecule with applications to information and biomolecular technology[17].

A QEMS device operates in a quantum domain if the quantum of energy is greater than thermal energy, $hf > k_B T$, and simultaneously has a quality factor above $10^5$. Here, $f$ is the frequency of the mechanical oscillator, $h$ is Planck's constant, $k_B$ is Boltzman's constant, and $T$ the temperature. Operating a device in this regime requires temperatures of the order of one Kelvin and oscillator frequencies greater than 500 MHz. Recent experimental progress indicates that these limits are achievable[18].

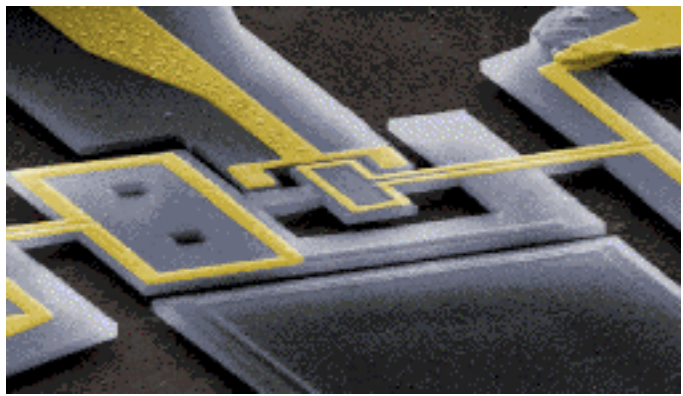

*Nano-mechanical torsion electrometer. (Courtesy M. Roukes, Caltech)*

Currently, progress towards QEMS is limited by the transducers of mechanical motion, including optical and electronic devices. In the optical readout scheme a cleaved fibre tip is brought near to the surface of the oscillator to form an optical cavity. Laser light reflected back from the surface of the oscillator undergoes phase modulation as the surface moves. Optical schemes do not scale well for the submicron scale oscillators required for QEMS, and hold no promise for motion transducers.

In the electronic schemes, wires patterned on the surface of the oscillator develops an induced EMF as the oscillator moves in an external static magnetic field. Sensitive mesoscopic electronic systems, such as the radio frequency single electron transistor (RF SET), may then be used to acquire an electronic signal of the oscillation. Currently the electronic readout schemes have not been able to achieve the required sensitivity to enable detection of motion at the quantum limit. As oscillator frequencies move to GHz, there is a limit to the time scales required for moving charge around on a small circuit.

**Single Spin Magnetic Resonance Force Microscopy:**

The ability to detect a single spin at a spatially resolved location would have considerable implications for data storage, quantum metrology, and quantum computing. A number of proposals for solid state quantum computers require the ability to detect the state of a single spin (eg the Kane[25] scheme). Magnetic resonance imaging (MRI) has been a significant technology for many decades and could even be called a quantum technology, based as it is on the intrinsic quantum degree of freedom called spin. The principal on which MRI is based is very simple: a small magnetic moment precessing in a magnetic field will emit a radio frequency field. The magnetic moment could be an electron or nuclear spin in a molecule or atom. If the precession frequency is driven on resonance by an external AC magnetic field, the resulting signal reveals something about the spin state of the molecule or atom and the external magnetic environment in which it is paced. Typically however it is impossible to detect a single spin and the signals obtained represent an ensemble average over a very large number of systems. A magnetic moment in an inhomogeneous magnetic field will experience a force. If this force can be detected and measured it can convey significant information about the source of the magnetic moment. The magnetic resonance force microscope (MRFM)[19] combines the imaging ability of MRI and spatial resolution of scanning force microscopy. By using QEMS devices considerable improvements in sensitivity (minimum force detectable) and spatial resolution can be hoped for. In the end, the objective is to image and measure the state of a single spin. This will require devices with atto-Newton-force sensitivity and mechanical resonance frequencies approaching the NMR Lamour frequency (10–100 MHz) and quality factors great than 10,000. At these frequencies and quality factors, and at the low temperatures at which such devices must operate, quantum mechanical noise in the mechanical oscillator will ultimately limit the performance of the device.

*Coherent quantum electronics.*

In a typical microfabricated electronic device, current is carried by either electrons or holes through incoherent transport. This means that a typical carrier experiences many inelastic collisions with other carriers and the bulk metal or semiconductor material. The observed current is an averaged process of drift, under an applied bias, against a background of erratic and highly disordered motion. Under such circumstances coherent quantum motion cannot occur (although quantum mechanics does determine macroscopic properties of super-conductance). In the last decade or so new devices have been fabricated which are so small and perfect that at, low temperatures, transport is dominated by elastic scattering processes. This regime is known as mesoscopic electronics[20]. It is a regime where the familiar rules of electronics, based as they are on a very classical description, no longer apply. In this realm, for example, we find that Ohm's law fails and conductance can be quantised in units of the quantum of resistance, $e^2/2h$, where $e$ is the electronic charge and $h$ is Planck's constant[15].

Many of the breakthroughs in mesoscopic electronics have come by finding ways to restrict the number of dimensions in which charge carriers can move. In a two-dimensional degenerate electron gas (2DEG) an electron is constrained to move in a plane with very tight confinement in one direction. Typically a 2DEG is formed at the atomically flat interface between two different semiconductor materials (such as GaAs and AlGaAs). In a magnetic field, motion of charge carriers in a 2DEG can exhibit highly quantum behaviour, such as the quantum Hall effect, which can only be explained by abandoning the notion of an independent charge-carrying particle and resorting to quantum many-body theory. When motion is restricted in two dimensions, a quantum wire is formed. Quantum wires can exhibit entirely new electronic properties such quantised conductance. Finally when motion is restricted in all three dimensions a quantum dot is formed. An electron or hole confined in all three dimensions has a discrete set of energies and is very much like an artificial atom[21]. There are a great variety of quantum dots, examples of which are: surface-gate defined trapping regions in a 2DEG containing thousands of electrons[22], self-assembled semiconductor quantum dots (formed at the interface between different semiconductors)[23] or chemically synthesised nanocrystals[24] with trapped excitons (a bound electron-hole pair), and a bound single charge on a dopant atom in a semiconductor[25], to name but a few.

**Superconducting quantum circuits.**
Superconducting Quantum Interference Devices (SQUIDs) are perhaps the best-known example of coherent quantum electronic devices. It is a superconducting ring that includes one or two Josephson junctions. A Josephson junction is an insulating barrier for current flow between two superconductors. The SQUID is a uniquely superconducting electronic device discovered in the mid-1960s that is most sensitive detector of magnetic fields. Commercial SQUIDs can measure magnetic fields 100 billion times smaller than the earth's magnetic field (which is very tiny to start with). With such sensors, signals as weak as the magnetic signature of electrical currents flowing in the human brain can be readily detected [Weinstock 96]. The Nobel-Prize- winning discovery of high temperature superconductors in the 1980s, which require only cheap liquid nitrogen rather than expensive liquid helium to cool, have paved the wave for many more practical applications. The high-temperature superconductors are brittle ceramics, which can be difficult to make into useful structures. However, intense investigation has led to the development of solid-state superconducting components and devices on a chip, using thin-film deposition technology and the same optical and electron-beam lithographic processing techniques that are at the core of the semiconductor industry. Compared to some of the other ideas in coherent quantum electronics, discussed above, the SQUID technology is already well developed. Still, there is much to be done, including the possible application to quantum computers. However, there is a serious roadblock that has yet to be overcome. A SQUID is normally found in a robust macroscopic quantum state with some phase, say 0 or  . However, to make a quantum computer it is required that the SQUID be put in a superposition state that has both phases 0 and   simultaneously. This is the signature of the elementary quantum logic bit or qubit, and it has just recently been observed in SQUIDs after years of effort. The problem was that the SQUID is a macroscopic quantum state that is not robust against certain types of decohering interactions with the environment. Any attempt to prepare a superposition state of 0 and   always ends up with the SQUID rapidly collapsing into either 0 or  . A recent variant on superconducting loops is the single-Cooper-pair box (SCPB). In such a device, a small superconducting island is coupled by weak tunnel junctions to a Cooper-pair reservoir. The island is so small, and the tunneling sufficiently weak, that the Coulomb energy cost of adding one extra Cooper pair to the island is large enough to force a strict turn-stile-like motion of individual charges. It is now possible to use charge as a quantum degree of freedom and to create superpostions of distinct charge states on the CPB. Charge and phase are canonically conjugate variables, so there is a nice complementarity in switching between phase qubits or charge qubits. . In a recent experiment, Nakamura has seen for the first time precisely this type of superposition in a nanofabricated SQUID-based device[26]. In a more recent experiment[27] the actual quantum degree of freedom used was a hybrid mix of charge and phase. Continued progress along these lines of decoherence control in SQUIDs could lead to the first scaleable quantum computer, although coupling distinct superconducting qubits together still presents a major challenge.

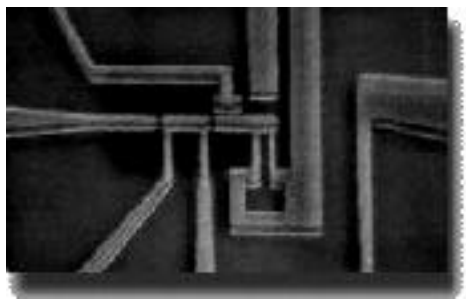

*Superconducting charge qubit prototype. (Courtesy Microdevices Lab, JPL.)*

**Quantum photonics.**

When light of the appropriate frequency is directed onto a semiconductor an electron-hole pair can be created, bound loosely by the coulomb interaction. Such as pair is called an exciton. If the electron and hole are additionally confined in a quantum dot the exciton can behave very much like an artificial atom, exhibiting a complex spectra[28]. Bound excitons can also be created electrically by injecting electrons and holes into a quantum dot[29]. Quantum dot excitons enable the world of coherent quantum electronics to talk to the world of photons and quantum optics, and provide a path to quantum opto-electronics. Much of the activity in this field is currently directed at developing a single photon source, which produces pulses of light with each pulse containing one and only one photon. Such a technology will enable secure quantum key distribution[30] and quantum optical computing. Currently the most promising paths to building a technology on excitonic quantum dots is based on self assembled interface dots and semiconductor nanocrystals. However the field is barely five years old.

**Spintronics.**

The quantum transport properties of charge carriers, such as an electron or hole, are not only determined by the quantum motion of the particle as a whole, but can involve internal degrees of freedom such as spin. Indeed the field of spin-dependent transport, or spintronics, is the fastest growing area of quantum electronics due largely to the possible applications to conventional information processing and storage[31]. The possible application to quantum computing of course has been noticed by many[32].

**Molecular coherent quantum electronics.**

Molecular electronics is the end point of the quest for smaller electronic devices.[33,34] The feasibility of using molecules as information storage and processing elements for classical computation was suggested long ago by Aviram and Ratner[35], but is only now having an impact on computer design, largely due to improvements in chemical synthesis. Both Hewlett Packard and IBM have research programs in molecular electronics[36]. The implementation of computing with molecular electronics is forcing a reappraisal of computer architectures[37]. Ultimately, molecular electronics may be able to utilise the self assembly that characterises biological systems, paving the way for a nanotechnology that mimics biological systems, a bio-mimetic nanotechnology. New materials such as carbon nanotubes and fullerenes provide a clear opportunity to engineer quantum nanodevices on a molecular scale, operating on either classical or quantum principles. The importance of this field was recognised by *Science* which nominated nanotube electronics as the most significant breakthrough of 2001[38]. While much of the current interest in molecular electronics is not based on coherent quantum transport, this is beginning to change. Recently a electronic interferometer was demonstrated, which used on a carbon nanotube[39]. In a more recent experiment still we can begin to see the merging of QEMS technology and molecular electronics. In this experiment charge was conducted trough a $C_{60}$ molecule which formed an island of a single electron transistor. However the molecule could vibrate mechanically which modulated the conductance curves in such a way that the Terahertz vibration frequency of the molecule could be observed[40]. It is not hard to imagine how such devices could approach a mesoscopic version of ion trap quantum computing experiments.

**Solid-State Quantum Computers.**
While much of the interesting mesoscopic electronics is driven by the need to make smaller conventional computing devices, the possible application to quantum computing is obvious and compelling due to the potential robustness and scalability of solid state devices. Some of the earliest suggestions for quantum computing were based on quantum dots[41]. There are now a very large number of schemes, some of which are being pursued in the laboratory[42]. As mentioned above there has been considerable progress in harnessing superconducting quantum degrees of freedom for quantum information processing, and the first all solid-state qubit has been demonstrated in such a device[27]. A solid state quantum computer is probably the most daunting quantum technological challenge of all and will require huge advances in almost all the areas of quantum technology we have discussed.

## *Quantum optics.*

Quantum optics is many things to many people, but at the core is the recognition that light is not a classical wave, but rather a quantum entity with both wave and particle aspects[43]. The additional particle character of light gives an additional quantum knob to turn in the engineering of photonic systems. In fact it is now possible to, in a sense, program digital, nonlocal correlations in collections of photons to improve the sensitivity of optical systems by many orders of magnitude.

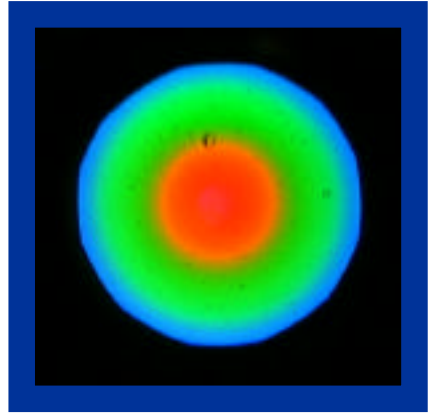

*Entangled photon pair production via parametric downconversion. (Courtesy of Quantum Internet Testbed, JPL.)*

There are many applications to sensors and quantum optical devices of unparalleled sensitivity. The breakthroughs and advances in quantum cryptography, and cavity QED quantum computing, spring from this newfound ability to engineer and design novel quantum states of light. In the last five years, there has been an explosion of research into the employment and utilization of quantum entangled photonic systems for scientific and technological advantage. A recent article by in *Physics Today* illustrates how quantum correlations have just been used to completely revolutionize the field of *Optical Metrology*[44].

Quantum optics has long provided the ideal experimental context to investigate some of more puzzling predictions of quantum mechanics[45]. There are two reasons for this. Firstly, at optical frequencies the world is very cold ($hf >> k_B T$), secondly, photons do not interact with each other, and only weakly with matter if resonance is avoided. It is thus possible to maintain quantum coherence for long times. These properties make photons ideal for transporting quantum information, but not so good for information processing. At least that was the traditional view. In 2000, Knill, Laflamme and Milburn discovered a way to implement quantum information processing using only linear optical devices[46]; linear optical quantum computation. This has opened up a new approach to quantum optical technology for metrology, control, communication and computation.

**Quantum Optical Interferometry:**

Recently, several researchers have been able to demonstrate theoretically that quantum photon entanglement has the potential to also revolutionize the entire field of optical Interferometry—by providing many orders of magnitude improvement in interferometer sensitivity[47]. The quantum entangled photon interferometer approach is very general and applies to many types of interferometers. In particular, without nonlocal entanglement, a generic classical interferometer has a statistical-sampling shot-noise limited sensitivity that scales like $1/\sqrt{N}$, where N is the number of particles (photons, electrons, atoms, neutrons) passing through the interferometer per unit time. However, if carefully prepared quantum correlations are engineered between the particles, then the interferometer sensitivity improves by a factor of square-root-of-N, to scale like 1/N, which is the limit imposed by *the Heisenberg Uncertainty Principle.* For optical (laser) interferometers operating at milliwatts of optical power, this quantum sensitivity boost corresponds to an eight-order-of-magnitude improvement of signal to noise. This effect can translate into a tremendous science pay-off for NASA missions. For example, one application of this new effect is to fiber optical gyroscopes for deep-space inertial guidance and tests of General Relativity (Gravity Probe B). Another application is to ground and orbiting optical interferometers for gravity wave detection, Laser Interferometer-Gravity Observatory (LIGO) and the European Laser Interferometer Space Antenna (LISA), respectively. Other applications are to Satellite-to-Satellite laser Interferometry (SSI) proposed for the next generation Gravity Recovery And Climate Experiment (GRACE II). In particular, quantum correlations employed in SSI could improve the orbital sensitivity to gravity enough to give unprecedented accuracy in measuring Earth gravitational anomalies from space with a resolution of 1 km or less. Such a sensitivity could be used for orbital oil prospecting or measuring water content of aquifers. The use of quantum correlated optical interferometers in NASA missions such as these would give a quantum breakthrough in sensitivity and performance.

**Quantum Lithography and Microscopy:**

In addition to these developments, it was recently shown that optical quantum entanglement effects can be used in *Quantum Interferometric Lithography*[48]. This application brings about a breakthrough in lithographic resolution by overcoming the diffraction limit—allowing features to be etched that are several factors smaller than the optical wavelength. This application has

sub-100 nm fabrication resolution at low cost. In addition to writing features much smaller than the wavelength, it is also possible to read back that information with this resolution. In classical optical microscopes, the finest detail that can be resolved in, say, a translucent micro-organism can be no smaller than the optical wavelength. Using the same entangled photon tricks as described above, one can image features substantially smaller than a wavelength. In both lithography and microscopy, the conventional classical road to finer resolution is to actually reduce the wavelength. But it is very hard (and expensive) to make imaging elements at UV and x-ray scales. In addition, such high frequency photons pack quite a punch and damage the object to be imaged or written on. Employing a new quantum knob to turn completely overcomes this bottleneck.

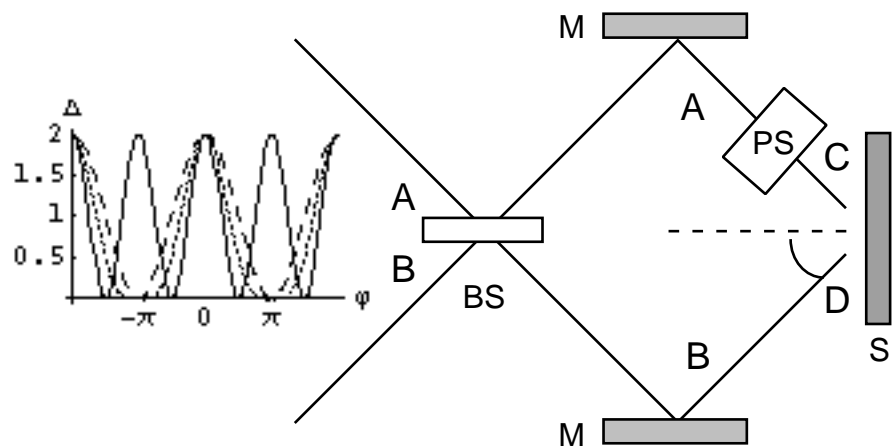

*Quantum lithography set up: Entangled photons can write finer lines that are twice as close together than possible classically.*

**Photon Squeezing:**

Optical squeezing is an analog noise-reduction process, similar to the idea of using entangled states to implement a quadratic scaling of signal-to-noise in optical Interferometry. The idea is to manipulate the bounds set by the *Heisenberg Uncertainty Principle* to squeeze (suppress) noise fluctuations in phase into noise fluctuations in number. Then, in an interferometer that is sensitive only to photon phase, you can gain a quadratic improvement in signal-to-noise. This concept was proposed in the 80's for improving the sensitivity of LIGO in particular and optical Interferometry in general. The application to LIGO appears to be elusive, but there has been interesting progress in using these states of light in supersensitive spectroscopy[49].

**Noninteractive Imaging:**
Using single photons in an interferometric setting, it is possible to image an object, without a photon actually interacting with the object. Hence, one can see what something looks like, without shining any light on it. One of the more bizarre predictions of quantum optics, this result has been born out in a series of elegant experiments by Paul Kwiat then at Los Alamos and Anton Zeilinger in Austria [Kwiat 96]. Some interesting applications would be the capability of imaging light-sensitive living cells without exposing them to light! What is even stranger, is the recent suggestion that quantum microscopy techniques can be combined with this noninteractive effect to perform subwavelength imaging of objects situated in total darkness[50].

**Quantum Teleportation:**
Quantum Teleportation is another nonintuitive idea that arose out of the field of *Quantum Information Theory*, discovered by several IBM researchers investigating the transmission of quantum information and quantum states[51]. From the *Heisenberg Uncertainty Principle*, it is known that it is not physically possible to copy a quantum state exactly. Any attempt to copy a state involves making a measurement to see what you want to duplicate, but this destroys the original state to be copied. However, counter-intuitively, it is possible to "teleport" a quantum state. That is, Alice can destroy a quantum state in Atlanta—only to have an exact duplicate appear instantaneously in Bob's lab in Boston, without Alice or Bob ever learning anything at all about this state. This phenomenon has been demonstrated experimentally by several groups[52].

*Coherent matter technology.*

The technological advances springing from quantum optics come from realizing that the classical wave-like nature of light is complemented by a particle picture, which gives a new degree of freedom to manipulate in terms of single photon correlations. In the case of atoms, a single neutral atom is usually treated quite well as a classical particle, say, a billiard-ball-like object. This description of atoms as point classical particles can be used to derive much of classical statistical mechanics, such as the *Ideal Gas Law*. Because of their large mass, compared to electrons, the wave nature of a single whole atom is a bit hard to tease out. The quantum atomic technological revolution has seen the capability of manipulating the coherent wave nature of atoms arrive in the lab in the past 10 years or so. Advances in laser cooling techniques[53] have given rise to a plethora of new phenomenon utilizing the wave nature of bulk atomic matter. Because of the large and quantized mass of atoms, Interferometric wave devices for atoms are uniquely sensitive to inertial and gravitational effects. This fact makes the technology a breakthrough advance in accelerometers, gyroscopy, gravity gradiometry, for applications to inertial guidance, geo-prospecting, tests of General Relativity, and the measurement of the geophysical structure of Earth from Space.

In the last decade it has become possible to trap and cool ensembles of atoms to the nanokelvin range. The most striking achievement of this field is the demonstration of Bose-Einstein condensation of a trapped gas of alkali atoms[54]. Recent developments have indicated the possibility of an atom laser: a source of coherent matter waves. In this aspect of the project It is possible atomic BECs will provide a path to high precision measurement[55] and experimental progress now makes this a real possibility with significant technological applications, particularly for gravimetry. New atom optical devices seek to trap and control atoms along magnetic wire guides fabricated at a micron scale on a substrate using technology borrowed from the microelectronics industry[56]. Such devices may provide a path to new high precision measurement and even quantum computing.

**Atom Optics:**
The advent of laser cooling techniques has created a new field of study analogous to the classical study of light optics, but here it is called atom optics. Atom optics allows us to treat the quantum wave nature of neutral atoms in a way that is parallel to the classical wave treatment of light. Using combinations of nanotechnology and laser cooling techniques, it is possible to make atom mirrors, atom lenses, atom diffraction gratings, atom fiber-optic wave guides, and atom traps[57]. In all cases, the devices guide, focus, diffract, or reflect atoms coherently, so that the phase of the atom wave function is preserved. Thus it is possible to make atom interferometers, atom lasers, atom trampolines, and other exotic matter-wave devices. Again, because of the relatively large mass of atoms, these devices are uniquely sensitive to gravitational and inertial effects[58]. In addition to the inertial sensing aspects, Mara Prentiss at Harvard has a program under way to use atom interferometers as a new type of lithographic device for nanofabrication[59].

**Quantum Atomic Gravity Gradiometry:**
For another application, recent experimental developments at Yale by Kasevich in the field of *Quantum Atomic Gravity Gradiometers (QuAGG),* have led to a new type of gravity gradiometer based on neutral-atom, matter-wave Interferometry[60]. These gradiometers are, in prototype, as sensitive as current superconducting state-the- art gradiometers, but without the requisite need for liquid helium coolants that currently limit satellite lifetime or the superconductors. In QuAGG, the wave-nature of the atoms is being manipulated by techniques that are the signpost of the second quantum revolution. Quantum correlation effects have the potential of magnifying the sensitivity of the QuAGG by many orders of magnitude, providing incomparable precision in gravity-field mapping of Earth or the other planets from space. These gradiometers also have applications to geological surveying and tests of *General Relativity*. A fledgling JPL effort to develop a QuAGG for space observation of the Earth's gravitational field are now underway.

**Atom Lasers:**
An atom laser is to an incoherent atomic beam as an optical laser is to a light bulb. The analogy is very good. In an optical laser the photons are all of one frequency and move in a well-defined direction all with the same phase. In the atom laser the atoms all have the same velocity and phase, exploiting the coherent wave-like nature of matter that is dictated by quantum mechanics. Just as the real progress in quantum optics did not occur until the invention of the optical laser in the 60's, so too is this the beginning of the coherent matter revolution with the atom laser now at our disposal. One application is to lithography, with the capability of creating neutral atom, matter-wave interferograms and holograms in order to write 2D and 3D patterns at the atomic level[59] . Another application is to inertial and gravitational sensors, where the atom wave is extraordinarily sensitive—due to its large mass compared to the photon. (A photon has an effective mass given by its energy divided by the speed of light squared, as per Einstein. This is on the order of eV, where the mass of an atom can be many keV[47].) Employing atom laser sources in matter wave gravity gradiometers promises to boost the sensitivity of these devices by yet further orders of magnitude. Early results from the group of Ketterle at MIT[61], have since been followed by Bill Phillips at NIST[62], Gaithersberg, and Ted Hänsch[63] at the Max Planck Institute for Quantum Optics.

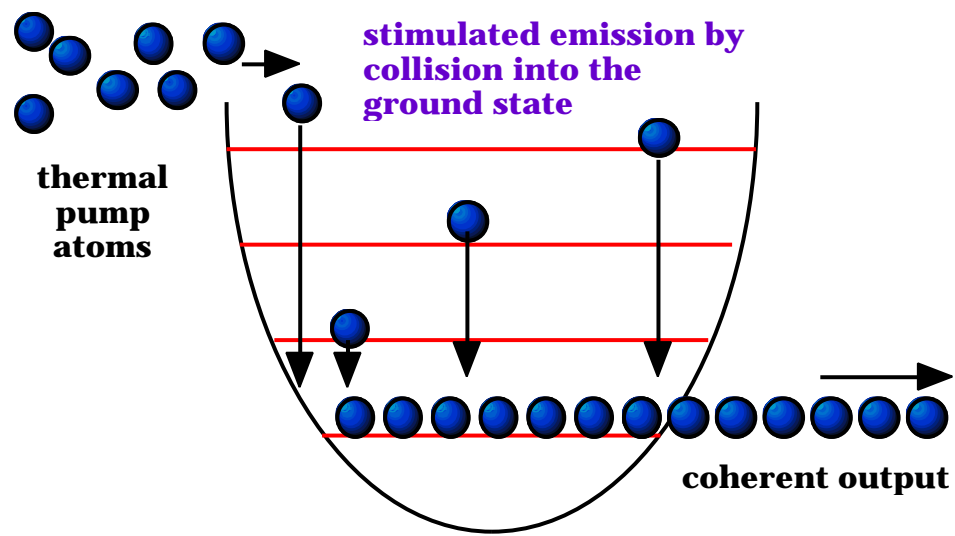

*Model of atom laser emitting a coherent beam of atom waves.*

## Conclusion.

The preceding survey may suggest that progress towards a future quantum technology is well in hand. However it is probably fairer to say that we have only begun to take the first halting steps down this path. Current efforts are still very much basic science and an enormous effort would be required to move even the most advanced of these research programs to development phase. Commercial applications are even further away for many of the technologies, although there are some promising beginnings (quantum cryptography for example) . As yet there is no dedicated effort in quantum technology as such, and the field itself has only just begun to define itself as a discipline. However we are convinced that as these experiments move form the lab to the market place the emergence of a new discipline of quantum technology or quantum engineering is almost certain, and it is not at all too early to ask where the wise money should be invested.

## Acknowledgements.

A portion of this work was carried out at the Jet Propulsion Laboratory, California Institute of Technology, under a contract with the National Aeronautics and Space Administration. One of the authors (JPD) would like to thank the Office of Naval Research, the Advanced Research and Development Activity, the National Security Agency, and the Defense Advanced Research Projects Agency for additional support.